\documentclass[12pt]{article}
\usepackage{latexsym}
\usepackage{graphicx}
\usepackage[usenames,dvipsnames]{color}

\def\be{\begin{equation}}
\def\ee{\end{equation}} 
\def\bea{\begin{eqnarray}}
\def\eea{\end{eqnarray}}

\def\e{\emph}

\def\nbn{$N$-body problem}

\def\case#1/#2{\textstyle\frac{#1}{#2}}

\def\e{\emph}

\begin{document}

\small\begin{center}

{\bf \Huge{Quantum without Quantum}}

\vspace{.3in}

{\bf \Large{Julian Barbour}}\footnote{julian.barbour@physics.ox.ac.uk}

\vspace{.1in}

College Farm, The Town, South Newington, Banbury, OX15 4JG, UK.

\end{center}

\abstract{\small In my contribution to the collection at https://dd70th.weebly.com marking the 70th birthday of David Deutsch I suggest that hitherto unrecognised properties of the Newton gravitational potential made scale-invariant through multiplication by the N-body root-mean-square length hint at redundancy of quantum wave functions for the explanation of physical effects.\normalsize}

\vspace{.2in}

{\bf My contribution.} Science is impossible without variety but may not need wave functions $\psi$. Particles in space separated by distances $r_{ij}=|{\bf r}_i-{\bf r}_j|, i, j=1,\dots N$, define a minimal model universe. Particles with masses $m_i,~\sum_im_i=1,$ have, respectively, root-mean-square and mean-harmonic lengths
\be
\ell_\textrm{\scriptsize rms}:=\sqrt{\sum_{i<j}m_im_jr_{ij}^2},~~\ell_\textrm{\scriptsize mhl}^{-1}:=\sum_{i<j}{m_im_j\over r_{ij}}.
\ee
Their ratio $\ell_\textrm{\scriptsize rms}/\ell_\textrm{\scriptsize mhl}=C$, called the complexity in [1], is the Newton potential made scale-invariant, and a sensitive measure of clustering; if the particles cluster, $\ell_\textrm{\scriptsize rms}$ changes little but $\ell_\textrm{\scriptsize mhl}$ decreases significantly. A long-standing problem in quantum gravity is the definition of time; many believe it should be a variable that describes the universe. In that spirit, $C$ is a candidate. Being positive definite, bounded below by a $C_\textrm{\scriptsize min}$, and unbounded above, $C$ defines a time that begins but never ends. Moreover, the universe is born maximally uniform, becoming thereafter evermore varied. Manuel Izquierdo [2] found typical shapes at birth and soon after for a 1000-particle universe in two dimensions (Figs.~1 and 2). They are critical points of $C$, and therefore central configurations (CCs) [3], but typical of many shapes with the same $C$ [4].

It is the simplicity of $C$'s definition and the structure of its CCs that leads me to question whether $h$ exists. The CC in Fig.~1, with $C$ at or very near $C_\textrm{\scriptsize min}$, is not new; ones much like it exist in three dimensions [5]. Newton's potential theorem explains their strikingly uniform density within an almost perfect sphere. The 2D rather than 3D calculation explains the slight radial density decrease in Figs.~1--3. The remarkable variety of the voids and filaments in Fig.~2 were a fortuitous discovery of Izquierdo, who, unlike previous researchers, looked for many-particle CCs with $C$ not only very close to $C_\textrm{\scriptsize min}$, but also a bit above it (by $\approx 1.5$\% in fact). 

{\includegraphics[width=0.6\textwidth]{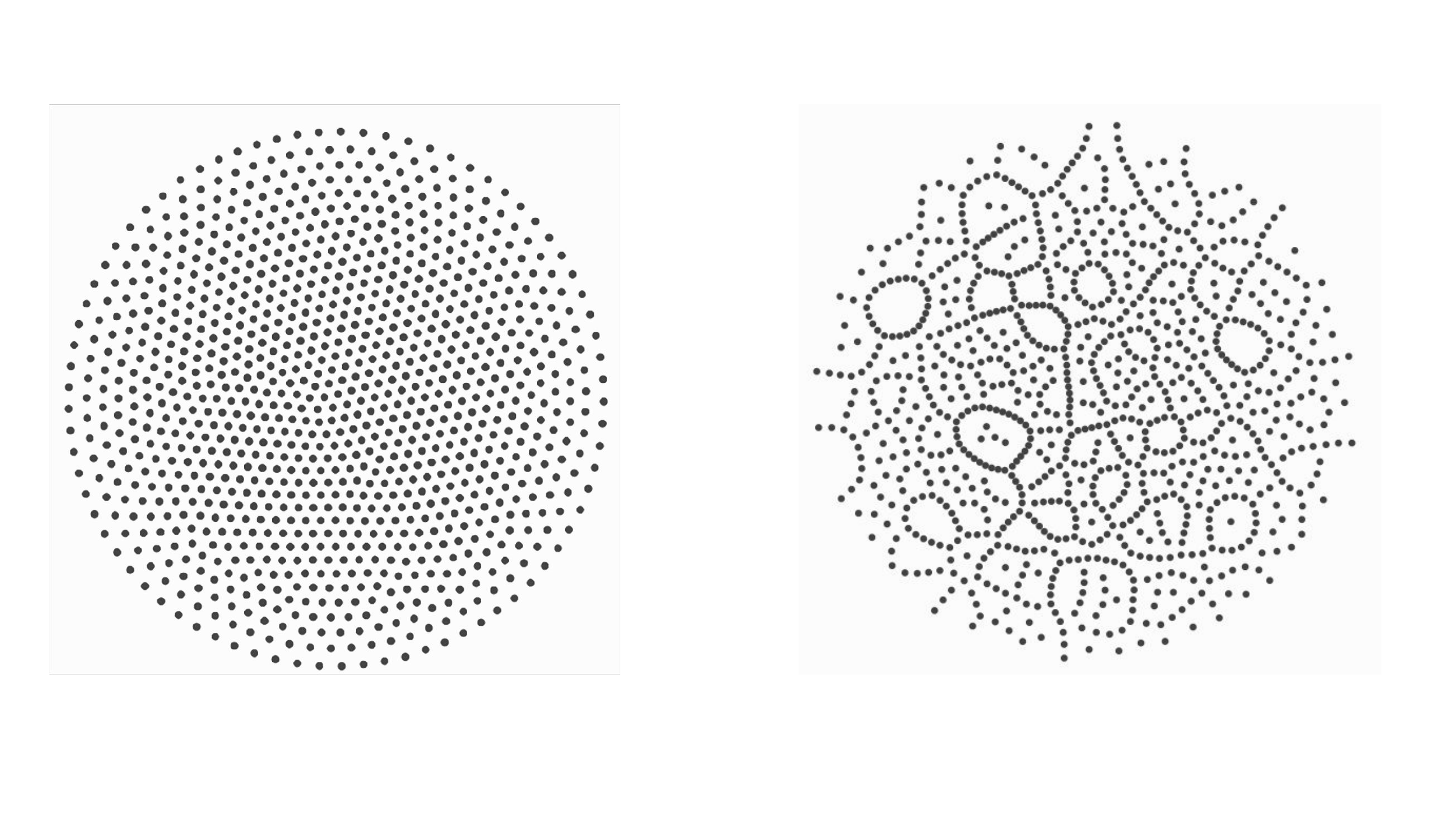}~~\includegraphics[width=0.38\textwidth]{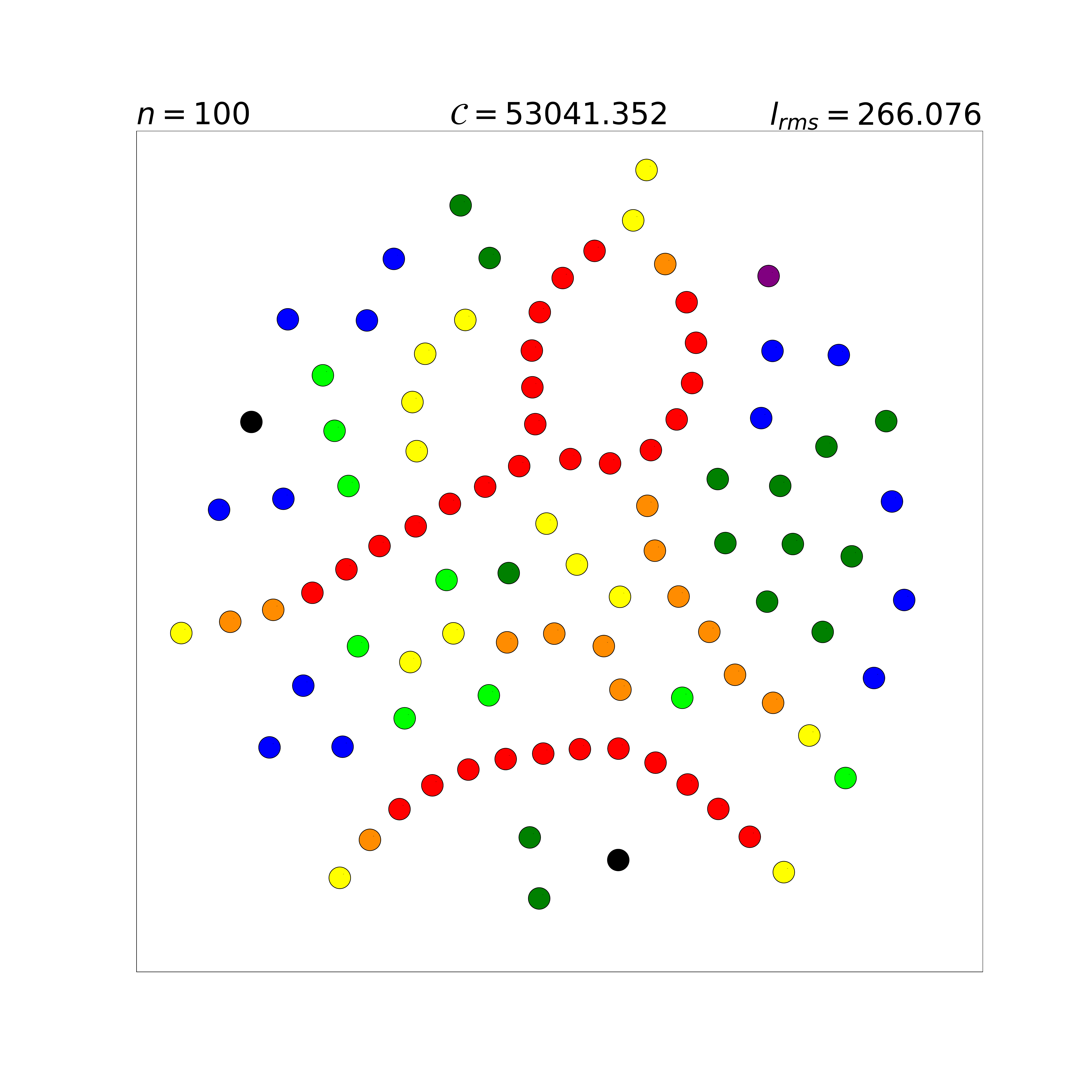}

~~~~~~~~~~~~Fig.~1~~~~~~~~~~~~~~~~~~~~~~~~~~~~~~~~Fig.~2~~~~~~~~~~~~~~~~~~~~~~~~~~~~~~~~~~~~~~~~Fig.~3

\vspace{.1in}

Now for my doubts, expressed briefly due to the nature of this note. The quasi-crystalline structure of Fig.~1 is the first. In fact, BCC crystallisation in CCs of 100,000 particles was found in [6]. Crystals, indeed all solids, are seen as quintessential quantum structures explained by wave-function antisymmetry of fermions [7]. I'm not aware anyone has suggested scale-invariant variety by itself can make a wave function $\psi$ redundant, but the simple fact that there are vastly more ways to increase $C$ from its value for Fig.~1 by changing all the $r_{ij}$ slightly than by moving a few closer together explains [4] why the smallest $r_{ij}$ in the filaments of Fig.~2 are very nearly equal. Remarkably, this effect is repeated in other filaments with successively greater separations but with no obvious statistical explanation. Izquierdo's red--orange--yellow coloured sequence of filaments in a typical 100-particle CC in Fig.~3 highlights the effect; it would be even more striking without the $r_{ij}$-increasing 2D edge effect. Is it fanciful to see in Figs.~2 and 3 protein chains in protozoic cells? If work in hand shows that, in all such CCs, the separations go up in steps that to a good accuracy are the same, that would cast more doubt on the need for wave functions. The paramount condition for that is met: correlations of arbitrarily long range exist everywhere in profusion. There are $\approx N$ of them between the $N(N-1)/2$ separations $r_{ij}$, which just $3N-6$ coordinates determine. Figure 2 is redolent of quantum-like probabilities; for example, given as a Bayesian prior that a closed loop of $x$ links exists somewhere, how many particles can one expect to find within it? It is the correlations between a source, two slits, and the pattern of fringes on an emulsion that are so hard to explain without interfering $\psi$ branches. The same is true of the Bell inequalities, now confirmed over seemingly arbitrary distances. Figures 1--3 suggest that the holistic nature of shapes with given $C$, whether CCs or not, may allow a top-down explanation of facts that baffle in a reductionist, bottom-up approach. In this connection, consider the actual evidence for quantum mechanics. In the early days it was mostly in photographs taken in laboratories. Indeed, what might be called generalised photographs---records encoded in distributions of matter---remain the sole source of evidence for wave functions. Examples are the huge detectors at the LHC. But they and it are the tiniest part of the universe. Were the theoreticians who invoked wave functions `looking in the wrong direction'. Did they need, but could not get, a snapshot of the whole universe containing within it the laboratory photographs? Suppose a dime, representing the region captured in such a photograph, is placed anywhere on the images in Figs.~2 or 3. The single condition that $C$ be critical fully explains the highly correlated structure the dime covers. There is no need for any $\psi$. Lack of space precludes further discussion here, but as Tim Koslowski said, if there is no wave function, that at least solves the measurement problem of how it collapses.

\flushleft{\bf Thanks.}} To Manuel Izquierdo for the figures; Richard Montgomery, Alain Chenciner, and Alain Albouy for teaching me much of what I know about the Newtonian \nbn; and my coauthors Tim Koslowski and Flavio Mercati of [1]. A great deal of what I discuss in [4] has arisen from discussions with them, above all Tim's fermionic interpretation of the particles and his insight, critical in [4], that, independently of any conjectured wave function $\Psi$ of the universe, a Born-type probability density exists on the isocomplexity subspaces of any shape space.

\flushleft{\bf References.}

\vspace{.1in}

1. J Barbour, T Koslowski, and F Mercati \e{Phys. Rev. Lett.} {\bf 113} 181101 (2014).\\ 
2. M Izquierdo, ``Filaments and voids in planar central configurations'' arXiv:2110.09855.\\
3. D Saari, ``Central configurations---a problem for the twenty-first century'' [PDF]psu.edu.\\
4. J Barbour, Online talk at https://www.youtube.com/watch?v=vskzeQc9XpI and podcast at https://youtu.be/I1mnQPdu44E with part relating to this note starting at 1hr32min.\\ 
5. R Battye, G Gibbons, and P Sutcliffe, ``Central configurations in three dimensions'' arXiv:hep-th/0201101v2.\\ 
6. Noted in [5] (their citation [29]).\\
7. H Brown, ``The reality of the wavefunction: arguments old and new'' [PDF]psu.edu.\\

\normalsize

\end{document}